\documentclass[12pt,preprint]{aastex}

\begin{document}

\title{Stability of the toroidal magnetic field in stellar radiation zones}

\author{Alfio Bonanno}
\affil{INAF, Osservatorio Astrofisico di Catania,
       Via S.Sofia 78, 95123 Catania, Italy \\
       INFN, Sezione di Catania, Via S.Sofia 72,
       95123 Catania, Italy}
\email{alfio.bonanno@inaf.it}

\author{Vadim Urpin}
\affil{INAF, Osservatorio Astrofisico di Catania,
       Via S.Sofia 78, 95123 Catania, Italy\\
       A.F.Ioffe Institute of Physics and Technology and Isaac Newton 
       Institute of Chile, Branch in St. Petersburg,
       194021 St. Petersburg, Russia}
\email{vadim.urpin@uv.es}

\begin{abstract}
Understanding the  stability of the magnetic field in radiation zones is of crucial 
importance for various processes in stellar interior like mixing, circulation and angular momentum transport.
The stability properties of a star containing a prominent
toroidal field in a radiation zone is investigated  by means of  
a linear stability analysis in the Boussinesq approximation
taking into account the effect of thermal conductivity. 
The growth rate of the instability is explicitly calculated and the 
effects of stable stratification and heat transport are discussed in detail.
It is argued that the stabilizing influence of gravity can never entirely 
suppress the instability caused by electric currents in radiation zones
although  the stable stratification can significantly decrease the growth 
rate of instability
\end{abstract}


\section{Introduction} 
Magnetic fields localized in stellar radiation zones can play an important role in many essential phenomena  like mixing,
angular momentum transport and formation of the solar tachocline, for instance.

It is rather uncertain which field can be present in the radiation zone.  
Observations provide upper limits on the allowed strength of the magnetic 
field in the radiation zone of the Sun. For instance, helioseismological 
measurements suggest an  upper limit $B < 4 \times 10^7$ G as an order of magnitude estimate
(see, e.g., Friedland \& Gruzinov 2004). The recently measured oblateness of 
the Sun (Codier \& Rozelot 2000) implies that the strength of the magnetic 
field is lower than $7 \times 10^6$ G. The observations of the splitting of 
solar oscillation frequencies provide the upper limit of $B < 3 \times 10^5$ 
G for a toroidal field near the base of the convective zone (Antia et al. 
2000). For other stars estimates are less certain and only theoretical 
upper limits can be derived.  

The possible origin of the field (if it exists) is also unclear. Likely, a 
hydromagnetic dynamo cannot operate in the radiation zone, where 
no strong flows are available to sustain a vigorous dynamo action. 
Perhaps, relic magnetic fields acquired by the star at the early stage of evolution can 
persist there. These type of fields could have formed, for instance because of 
differential rotation, which could have stretched a weak  diffuse 
primordial seed field (see, e.g., Dicke 1979) into a dominant toroidal field.
Due to a high conductivity, 
the ohmic decay is very slow and the decay time can exceed several billion 
years. Therefore, once formed, the large-scale relic field 
would survive in the radiation zone during the life-time of a star. 

The magnetic field, however, can evolve in a radiation zone not only due
to ohmic dissipation but also because of the development of various 
instabilities. For instance, in differentially rotating radiation zones, 
the magnetorotational instability can occur. Most likely, however, the 
magnetized radiative zones rotate rigidly and the stability of magnetic 
configurations is determined by various current-driven instabilities. 
Such instabilities are well studied in cylindrical geometry in the 
context of laboratory fusion research (see, e.g., Freidberg 1973, 
Goedboed 1971, Goedbloed \& Hagebeuk 1972). For example, the stability 
properties of a pure toroidal field $B_{\varphi}$ are determined by the 
parameter $\alpha = d \ln B_{\varphi}(s) / d \ln s$  where $s$ is the 
cylindrical radius. The field is unstable to axisymmetric perturbations if 
$\alpha  > 1$ and to non-axisymmetric perturbations if $\alpha> - 1/2$ (Tayler 
1973a,b, 1980). Note, however, that the addition of  a even relatively weak 
poloidal field alters substantially the stability properties of the magnetic
configuration. For example, if the poloidal field is uniform and relatively 
weak, the instability condition reads $\alpha > -1$  at variance with the 
condition of instability for a purely toroidal field (see, e.g., Knobloch 
1992; Dubrulle \& Knobloch 1993) which predicts that an unstable toroidal 
field configuration has $\alpha > -1/2$. In astrophysical conditions, the 
instability caused by electric currents might have various characteristic properties
(see Bonanno \& Urpin 2008a,b).
In the presence of both azimuthal and axial  fields, non-axisymmetric disturbances with large azimuthal wavenumbers $m$ 
turn out to be most rapidly growing. Unstable disturbances exhibit a resonant 
character, i.e. the wave vector $\vec{k} = (m/s) \vec{e}_\varphi + 
k_z \vec{e}_z$ approximately satisfies the condition of magnetic resonance, 
$\vec{B} \cdot \vec{k} = 0$ where $k_z$ is the wavevector in the axial 
direction and  $\vec{B}$ is the magnetic field.  The length scale of this 
instability depends on the ratio of poloidal and azimuthal field components 
and it can be very short, while the width of the resonance turns out to be 
very narrow. For this reason, its excitation in direct numerical simulations can be 
problematic.

Stability of the spherical magnetic configurations is studied in much less 
detail and even the overall  stability properties of radiation zones are rather unclear.
Braithwaite \& Nordlund (2006) have studied the 
stability of a random initial field in the stellar radiative zone by means of direct numerical simulations
and found that the stable magnetic configurations generally have the form of tori with 
comparable poloidal and toroidal field strengths. Numerical modeling by 
Braithwaite (2006) confirmed that the toroidal field with $B_{\varphi} \propto 
s$ or $\propto s^2$ is unstable to the $m=1$ mode as it was predicted by 
Tayler (1973). However, even a purely toroidal field can be stable in the 
region where it decreases rapidly with $s$. Note that a purely toroidal field 
cannot be stable throughout the whole star because the stability condition 
for axisymmetric modes ($\alpha < 1$) is incompatible with the condition 
that the electric current in the $z$-direction has no singularity at $s 
\rightarrow 0$ which implies $\alpha > 1$. The stability of the toroidal 
field in rotating stars has been considered by Kitchatinov (2008) and 
Kitchatinov \& R\"udiger (2008) who argued that the magnetic instability is 
essentially three-dimensional and determined the threshold field strength 
at which the instability sets. Estimating this threshold in the solar 
radiation zone, the authors impose the upper limit on the magnetic field  
$\approx 600$ G.

In this paper,  we consider the stability of the toroidal field in radiation 
zones by taking into account stratification and thermal conductivity. It 
turns out that magnetic configurations  can be stable or unstable depending 
on the radial profile of the toroidal field. We argue that stable 
stratification can substantially decrease the growth rate of the instability 
but cannot suppress it entirely.   

%%%%%%%%%%%%%%%%%%% section 1
\section{Basic equations}
Consider the stability of an axisymmetric toroidal magnetic field in 
the radiation zone using a high conductivity limit. We work in spherical 
coordinates ($r$, $\theta$, $\varphi$) with the unit vectors ($\vec{e}_{r}$, 
$\vec{e}_{\theta}$, $\vec{e}_{\varphi}$). We assume that the toroidal field 
depends on $r$ and $\theta$, $B_{\varphi}= B_{\varphi}(r, \theta)$. 
In the incompressible limit, the MHD equations read  
\begin{eqnarray}
\frac{\partial \vec{v}}{\partial t} + (\vec{v} \cdot \nabla) \vec{v} = 
- \frac{\nabla p}{\rho} + \vec{g} 
+ \frac{1}{4 \pi \rho} (\nabla \times \vec{B}) \times \vec{B}, 
\end{eqnarray}
\begin{equation}
\frac{\partial \vec{B}}{\partial t} - \nabla \times (\vec{v} \times \vec{B}) 
= 0,
\end{equation}
\begin{equation}
\nabla \cdot \vec{v} = 0, \;\;\; \nabla \cdot \vec{B} = 0, 
\end{equation}
where $\vec{g}$ is gravity. In the basic state, the gas is assumed to be 
in hydrostatic equilibrium, then
\begin{equation}
\frac{\nabla p}{\rho} = \vec{g} + \frac{1}{4 \pi \rho} 
(\nabla \times \vec{B}) \times \vec{B}.
\end{equation}
We assume that the magnetic energy is subthermal and therefore $\vec{g}$ is
approximately radial. The equation of thermal balance reads in the 
Boussinesq approximation
\begin{equation}
\frac{\partial T}{\partial t} + \vec{v} \cdot (\nabla T - \nabla_{ad}T) =
\nabla \cdot (\kappa \nabla T),
\end{equation} 
where $\kappa$ is the thermal diffusivity and $\nabla_{ad} T$ is the adiabatic 
temperature gradient.

We consider a linear stability. Small perturbations will be indicated by 
subscript 1, while unperturbed quantities will have no subscript. Linearizing 
Eqs.(1)-(3) and (5), we take into account that small perturbations of the 
density and temperature in the Boussinesq approximation are related by
$\rho_1/\rho = -\beta (T_1/T)$ where $\beta$ is the thermal expansion 
coefficient. For small perturbations, we use a local approximation in the 
$\theta$-direction and assume that their dependence on $\theta$ is 
proportional to $\exp( - i l \theta)$, where $l \gg 1$ is the longitudinal 
wavenumber. Since the basic state is stationary and axisymmetric, the 
dependence of perturbations on $t$ and $\varphi$ can be taken in the 
exponential form as well. Then, perturbations are proportional to 
$\exp{(\sigma t - i l \theta - i m \varphi)}$ where $m$ is the azimuthal 
wavenumber. The corresponding wavevectors are $k_{\theta}=l/r$ and 
$k_{\varphi}=m/r \sin \theta$, respectively. The dependence on $r$ should be 
determined from Eqs.(1)-(3), (5). For the sake of simplicity, we assume 
that unperturbed $\rho$ and $T$ are approximately homogeneous in the 
radiation zone. This assumption does not change the main conclusions 
qualitatively but substantially simplifies calculations. 
We also neglect the effect of rotation. As we know from the work
of Pitts and Tayler (1986), rotation itself cannot remove instabilities of the
interchange type, although it can affect the shape the unstable modes.
For this reasons we argue that the conclusions 
of our investigation cannot be altered by the inclusion of rotation.

Eliminating all 
variables in favor of $v_{1r}$ and $T_1$, we obtain the following set of two coupled 
equations
\begin{eqnarray}
(\sigma^2 + \omega_A^2) v_{1r}'' +
\left( \frac{4}{r} \sigma^2 + \frac{2}{H} \omega_A^2 \right) 
v_{1r}' +  \left[ \frac{2}{r^2} \sigma^2
- k_{\perp}^2 (\sigma^2 +  \omega_A^2 ) \right.  
\\ \nonumber
\left. + \frac{2}{r} \omega_A^2 \left( \frac{1}{H} 
\frac{k_{\perp}^2}{k_{\varphi}^2} - \frac{2}{r} 
\frac{k_{\theta}^2}{k_{\varphi}^2} \frac{\sigma^2}{\sigma^2 + \omega_A^2}   
\right) \right] v_{1r} = - k_{\perp}^2 \beta g \sigma \frac{T_1}{T},
\\ 
\frac{\kappa}{r^2} \frac{\partial}{\partial r} \left[ r^2
\frac{\partial}{\partial r}
\left( \frac{T_1}{T} \right) \right]
- (\sigma + \kappa k_{\perp}^2) \frac{T_1}{T} = \frac{\omega_{BV}^2}{\beta g}
v_{1r},
\end{eqnarray}
where the prime denotes a derivative with respect to $r$ and
\begin{eqnarray}
\omega_A^2 = \frac{k_{\varphi}^2 B_{\varphi}^2}{4 \pi \rho}, \;\;\;
\omega_{BV}^2 = - \frac{g \beta}{T} (\nabla_{ad} T - \nabla T)_r, 
\\ \nonumber
k_{\perp}^2 = k_{\theta}^2 + k_{\varphi}^2, \;\;\;
\frac{1}{H} = \frac{\partial}{\partial r} (r B_{\varphi}). 
\end{eqnarray}

Some general stability properties can be derived directly from Eqs.(6)-(7).
Consider perturbations with a very short radial wavelength for which one
can use a local approximation in the radial direction, such as $v_{1r} 
\propto \exp( -ik_{r} r)$, where $k_{r}$ is the radial wavevector. If $k_{r} 
\gg \max(k_{\theta}, k_{\varphi})$, then Eqs.(6)-(7) can be reduced with 
the accuracy in terms of the lowest order in $(k_{r} r)^{-1}$ to 
the following set 
\begin{equation}
-(\sigma + \kappa k^2) \frac{T_1}{T} = \frac{\omega_{BV}^2}{\beta g} v_{1r}, 
\;\;\;\;\;
k_{r}^2 (\sigma^2 + \omega_A^2) v_{1r} = k_{\perp}^2 \beta g \sigma 
\frac{T_1}{T}, 
\end{equation} 
where $k^2 = k_{r}^2 + k_{\perp}^2$. The corresponding dispersion relation
reads 
\begin{equation}
\sigma^3 + \kappa k^2 \sigma^2 + \left( \omega_A^2 + \frac{k_{\perp}^2}{k^2}
\omega_{BV}^2 \right) \sigma + \kappa k^2 \omega_A^2 = 0.
\end{equation}
The conditions that at least one of the roots has a positive real part
(unstable mode) is determined by the Routh criterion (see, e.g., Aleksandrov,
Kolmogorov, \& Laurentiev 1985). For a particular case of a cubic equation,
these conditions can be easily obtained, for example, from expressions
derived by Urpin \& R\"{u}diger (2005). Since the quantities $\kappa$ and 
$\omega_A^2$ are positively defined, the only non-trivial condition 
of instability is $\omega_{BV}^2 < 0$ that is not satisfied in the 
radiation zone by definition. Therefore, modes a with short radial 
wavelength are always stable to the current-driven instability contrary to 
the conclusion obtained by Kichatinov (2008) and 
Kichatinov \& R\"{u}diger (2008). 

\section{Numerical results}

We assume that the radiation zone is located at $R_i \geq r \geq R$. The 
toroidal field can be represented as
\begin{equation}
B_{\varphi} = B_0 \psi(x) \sin \theta,
\end{equation}
where $x= r/R$ and $B_0$ is the characteristic field strength; $\psi \sim 1$ 
is a function of the radius alone. The dependence of $\psi$ on $x$ is 
uncertain in in the radiation zone and, in this work, we consider 
only the case where $\psi$ increases with $x$.  Other possibilities will be 
considered in a forthcoming paper. We parametrize $\psi(x)$ with  
the  following dependence 
\begin{equation}
\psi(x) = \Big (D_1 \exp \frac{x - 1}{d}  + D_2 \Big)  
\end{equation}
where $d$ characterizes the width of a field distribution and $D_1$ and $D_2$ 
are chosen so that $\psi(x=1)=1$ and $\psi(x=x_i)=0$ as shown in Fig.1; $x_i
=R_i/R$. We choose $x_i$ to be equal to $0.01$ from computational reasons. We 
have verified that our results are basically insensitive to the precise value 
of $x_i$ as long as it is close to the center. The situation where the 
field reaches its maximum at the outer boundary can model, for example, the 
radiative interior of a star with a convective envelope. In this case, the 
bottom of the convection zone is the location of the toroidal field 
generated by a dynamo action which can penetrate into the radiation zone.
This situation can also mimic the toroidal field in the liquid core of neutron 
stars. Likely, the magnetic field of these objects is generated by turbulent 
dynamo during the very early phase of the evolution when the neutron star 
is subject to hydrodynamical instabilities (see, e.g., Bonanno et al. 2003). 
Dynamo induced by turbulent motions generates the magnetic field of 
complex topology including small scale fields (Urpin \& Gil 2004). 
Large-scale dynamo is most efficient in the surface layers where the density 
gradient is maximal. Therefore, the generated field increases outward and 
reaches its maximum in the outer layers (Bonanno et al. 2005, 2006). This 
magnetic field  can be subject to current-driven instabilities after the end
of the unstable phase.  

Introducing the dimensionless quantities
\begin{equation}
\omega_{A0}^2 = \frac{B_{0}^2}{4 \pi \rho R_2^2}, \;\;\; 
\Gamma = \frac{\sigma}{\omega_{A0}}, \;\;\;
\delta^2 = \frac{\omega_{BV}^2}{\omega_{A0}^2}, \;\;\; 
\varepsilon = \frac{\omega_T}{\omega_{A0}}, 
\end{equation} 
where $\omega_T = \kappa/R_2^2$, we can transform Eqs.(6)-(7) into a 
dimensionless form. These equations with the corresponding boundary
conditions describe the stability problem as a non-linear eigenvalue 
problem. Fortunately, the main qualitative features of this problem 
are not sensitive to the choice of boundary conditions. That is 
why we choose the simplest conditions and assume that $v_{1r}=
T_1 = 0$ at $r=R_i$ and $r=R$. Note that the parameter $\delta$ is
large in radiation zones but, most likely, $\varepsilon$ is relatively
small if the magnetic field is not very weak. In calculations, we suppose
$\delta$ and $\varepsilon$ to be constant through the radiation zone. 

To illustrate the stabilizing effect of stratification on the instability,
we plot in Fig.2 the growth rate as a function of the stratification
parameter $\delta$ for several eigenmodes with a different number of nodes
in the radial direction. In these calculations, we neglect the thermal
conductivity, therefore, the stabilizing effect of gravity is most pronounced. 
The growth rate is always  maximal for the fundamental eigenmode ($n=0$) but
rapidly decreases with an increasing number of nodes $n$. Note that such 
a rapid decrease of $\Gamma$ with the number of an eigenmode is 
typical also for the case of neutral stratification $\delta=0$ (or no gravity 
$g=0$). The conclusion that $\Gamma$ decreases rapidly with $n$ confirms our
result obtained from Eqs.(6)-(7) in a short wavelength approximation (see
Section 2) that modes with a short radial lengthscale (large $n$) should be 
stable. This conclusion is at variance  with the statement by Kitchatinov \& 
R\"{u}diger (2008) that the most rapidly growing modes correspond to $n \sim
10^3$. According to our calculations, the fundamental mode turns out to be 
most rapidly growing. However, even the instability of this mode is entirely 
suppressed if $\delta \gtrsim9$. 
%\begin{figure}
%\includegraphics[width=8.5cm]{tlet3.eps}
%\caption{The dependence of the toroidal field on the spherical radius
%for three possible models.}
%%\label{angle}
%\end{figure}
%\begin{figure}
%\includegraphics[width=8.5cm]{Figure3.eps}
%\caption{The dimensionless growth rate as a function of $\delta^2$ for the
%fundamental eigenmode with $m=1$, $l=20$, $d=0.1$, and three values of
%$\varepsilon$: $10^{-2}$ (solid line), $2 \times 10^{-2}$ (dashed), and
%$4 \times 10^{-2}$ (dot-dashed).}
%\label{angle}
%\end{figure}
The effect of a thermal conductivity can change the properties of a
current-driven instability qualitatively. In Fig.3, we plot the dependence
of the growth rate on the parameter stratification $\delta^2$ for three
different values of the thermal conductivity, corresponding to $\varepsilon=
10^{-2}$, $2 \times 10^{-2}$, and $4 \times 10^{-2}$. The behaviour of all
curves is qualitatively similar: the growth rate is $\approx 1$ at
small $\delta$ decreases as
\begin{equation}
\Gamma \propto \delta^{-2} 
\end{equation}
for $\delta^2 > 100$ or, in  dimensional form,
\begin{equation}
\sigma \propto \omega_{A0} ( \omega_{A0}/\omega_{BV})^2.
\end{equation}
This dependence can be obtained also directly from the basic equations. The 
stabilizing effect of gravity in a linear theory is determined by the first 
term on the right hand side of Eq.(1). Linearization of this term yields 
$(\nabla p/ \rho)_1 \approx - \vec{g} \rho_1/\rho \approx \vec{g} T_1/T$ 
since $p_1 \approx 0$ in the Boussinesq approximation. Perturbations of the 
temperature are determined by thermal balance Eq.(7). Near 
the threshold of the instability when $\sigma$ is small, we have $T_1/T \approx 
- (\omega_{BV}^2 / \beta g \kappa k_{\perp}^2) v_{1r}$. Therefore, the
stabilizing contribution of gravity in Eq.(1) is of the order of 
$(\omega_{BV}^2 / \kappa k_{\perp}^2) v_{1r} \sim \delta^2 ( \omega_{A0}^2 /
\kappa k_{\perp}^2) v_{1r}$. On the other hand, the destabilizing effect of
electric currents in Eq.(1) is given by the Lorentz force. The order of
magnitude estimate of the Lorentz force yields $\sim B_{\varphi} B_{1 \varphi}
/ 4 \pi \rho r$. Since the toroidal field is inhomogeneous in the basic state,
perturbations of the toroidal field are produced basically by perturbations
of the radial velocity and $B_{1 \varphi} \sim (B_{\varphi}/ \sigma H) 
v_{1 r}$ where $H$ is the radial lengthscale of the magnetic field. Using 
this expression, we obtain that the Lorentz force is of the order of $\sim 
(\omega_{A0}^2 / a \sigma) v_{1r}$ where $a=H/R$. Equating the stabilizing 
contribution of gravity to the destabilizing contribution of the electric 
current, we obtain
\begin{equation}
\sigma \sim \omega_{A0} \; \frac{\varepsilon}{a \delta^2} \; (k_{\perp}^2 
R_2^2).
\end{equation}   
If $\varepsilon$ is fixed, this expression yields the dependence $\Gamma 
\propto \delta^{-2}$ that follows from numerical calculations.

The relation (14) implies that even a very strong stable stratification 
cannot entirely suppress the instability. The growth rate turns out to be
non-vanishing even for large $\delta$ at variance with 
the case where thermal conductivity is neglected (see Fig.2). 

\section{Conclusions}
We have considered the effects of stratification and thermal conductivity
on the stability of a toroidal field in stellar radiation zones. 
The magnetic configuration with a predominantly toroidal field can be 
formed, for example, due to differential rotation during the early stage 
of stellar evolution if the star has got even a weak seed field.

It turns out that stable stratification can suppress the current induced
instability of the toroidal field if perturbations are not influenced
by the thermal conductivity (very small $\varepsilon$). The instability 
does not arise if the Brunt-V\"{a}is\"{a}l\"{a} frequency is greater than 
$\sim 9 \omega_{A0}$. Since $\omega_{BV}$ is typically high in radiative zones
($\sim 10^{-3}-10^{-4}$ s$^{-1}$) the instability sets in only if the field
is very strong ($\geq 10^6-10^7$ G). Higher eigenmodes are suppressed 
stronger than the fundamental one and perturbations with short radial 
wavelength are always stable.

The thermal conductivity drastically changes the character of the instability.
This concerns particularly the behavior near the threshold of instability
where the growth rate is small. It turns out that the growth rate is 
non-vanishing for any stratification. Even very strong gravity cannot
suppress the instability entirely but it only decreases the growth rate. 
This sort of  behavior is typical for perturbations with any wavevector
$k_{\perp}$ and can be easily understood. A destabilizing effect of electric
currents in the momentum equation (1) originates from the last term on the
r.h.s. and is proportional to a perturbation of the magnetic field, $B_1$.
A perturbation $B_1$ is related to $v_1$ by $B_1 \propto v_1 / \sigma$ as
it follows from the linearized induction equation.  A stabilizing 
influence of gravity in the linearized momentum equation is proportional 
to a perturbation of the temperature, $T_1$. If the growth rate is small, 
then we can obtain from thermal balance equation (5) that $T_1 \propto
v_1 / \omega_T$. Comparing the destabilizing and stabilizing effects, we
see that stable stratification can never suppress the instability 
at variance with the $\kappa = 0$ case.

A decrease of $\sigma$ caused by stratification is inversely proportional 
to the Brunt-V\"ais\"al\"a frequency. If gravity is strong but the magnetic 
field is weak, the instability develops very slowly. Generally, for a 
sufficiently weak magnetic field, the growth rate can be comparable to
the inverse life-time of a star. It should be noted also that, most likely, 
the field does not decay to zero because of this instability. When the field 
becomes weaker, the growth rate of the instability decreases and the field 
cannot decay to values smaller than those resulting from a growth rate of 
the order of the inverse life-time of a star. Therefore, a weak field can 
generally be only slowly changed during the life of the star, although 
its radial profile can be unstable.

%\begin{figure}
%\plotone{f7.eps}
%\caption{The dependence of the growth rate and frequency on $x^{2} = 
%(ks)^{2}$ for radially decreasing $\Omega(s)$ ($q=4$) and for $\mu=
%\varepsilon=0.3$, $\delta=0.2$, $h=0.5$, $f_{s}=0.3$ and $f_{m}=0.2$.}
%\end{figure}

%To illustrate the effect of a relative orientation of the radial and
%azimuthal field, we plot in Fig.~8 the growth rate as a function of 
%$x^{2}$ for $\delta=-0.2$ (other parameters are same as in Fig.~7). 
%The new instability still may occur in the region of $x^{2}$ where the 
%magnetorotational instability is forbidden. The change of sign of 
%$\delta$ is not crucial for the growth rate which $\sim 0.05 \Omega_{e}$ 
%in this case as well. In the contrast to the case $\delta=0.2$, however, 
%only one couple of modes is unstable for $\delta=-0.2$. Note that the 
%change of sign of $\delta$ yields insignificant effect on the growth 
%rate of the magnetorotational instability at $x^{2}<15$.
%
%\begin{figure}
%\begin{center}
%\includegraphics[width=9.0cm]{cmri8.ps}
%\caption{Same as in Fig.~7 but for $\delta=-0.2$.}
%\end{center}
%\end{figure}
\vspace{0.5cm}

\noindent
VU thanks also INAF-Ossevatorio Astrofisico di Catania for hospitality and financial support.

%\begin{thebibliography}{99}
%
\begin{center}
{\large \bf  REFERENCES}
\end{center}
%\bibitem[]{}
\noindent
Aleksandrov, A.D., Kolmogorov, A.N., \& Laurentiev, M.A. 1985. Mathematics:
Its Content, Methods, and Meaning (Cambridge, MIT Press)\\[3mm]
%\bibitem[]{}
Antia, H.M., Chitre, S.M., Thompson, M.J. 2000. A\&A, 360, 335\\[3mm] 
%\bibitem[]{}
Bonanno, A., Rezzolla, L., \& Urpin, V. 2003. A\&A, 410, 33\\[3mm]
%
%\bibitem[]{}
Bonanno, A., Urpin, V., \& Belvedere, G. 2005. A\&A, 440, 199\\[3mm]
%
%\bibitem[]{}
Bonanno, A., Urpin, V., \& Belvedere, G. 2006. A\&A, 451, 1049\\[3mm]
%
%\bibitem[]{}
Bonanno, A., Urpin, V. 2008a. A\&A, 477, 35\\[3mm]
%
%\bibitem[]{}
Bonanno A., Urpin V. 2008b. A\&A, 488, 1\\[3mm]
%
%\bibitem[]{}
Braithwaite, J., Nordlund, A. 2006. A\&A, 450, 1077\\[3mm]
%
%\bibitem[]{}
Braithwaite, J. 2006. A\&A, 453, 687\\[3mm]
%
%\bibitem[]{}
Dicke, R.H. 1979. ApJ, 228, 898\\[3mm]
%
%\bibitem[]{}
Dubrulle, B., \& Knobloch, E. 1993, A\&A, 274, 667\\[3mm]
%
%\bibitem[]{}
Freidberg, J. 1970. Phys. Fluids, 13, 1812\\[3mm]
%
%\bibitem[]{}
Friedland, A., Gruzinov, A. 2004. ApJ, 601, 576\\[3mm]
%
%\bibitem[]{}
Goedbloed, J.P. 1971. Physica, 53, 501\\[3mm]
%
%\bibitem[]{}
Goedbloed, J.P., Hagebeuk, H.J. 1972. Phys. Fluids, 15, 1090\\[3mm] 
%
%\bibitem[]{}
Godier, S., Rozelot, J.-P. 2000. A\&A, 355, 365\\[3mm]
%
%\bibitem[]{}
Kitchatinov, L. 2008. Astron. Rep, 52, 247\\[3mm]
%
%\bibitem[]{}
Kitchatinov, L., R\"{u}diger, G. 2008, A\&A, 478, 1\\[3mm]
%
%\bibitem[]{}
Knobloch, E. 1992. MNRAS, 255, 25\\[3mm]
%
%\bibitem[]{}
Pitts, E.,  Tayler R.J., 1986, MNRAS 216, 139\\[3mm]
%
%\bibitem[]{}
Tayler, R. 1973a. MNRAS, 161, 365\\[3mm]
%
%\bibitem[]{}
Tayler, R. 1973b. MNRAS, 163, 77\\[3mm]
%
%\bibitem[]{}
Tayler, R. 1980. MNRAS, 191, 151\\[3mm]
%
%\bibitem[]{}
Urpin V., Gil J. 2004. A\&A, 415, 305\\[3mm]
%
%\bibitem[]{}
Urpin V., R\"{u}diger G. 2005. A\&A, 437, 23\\[3mm]
%
%\end{thebibliography}
%
\begin{figure}
\begin{center}
\includegraphics[width=8.5cm]{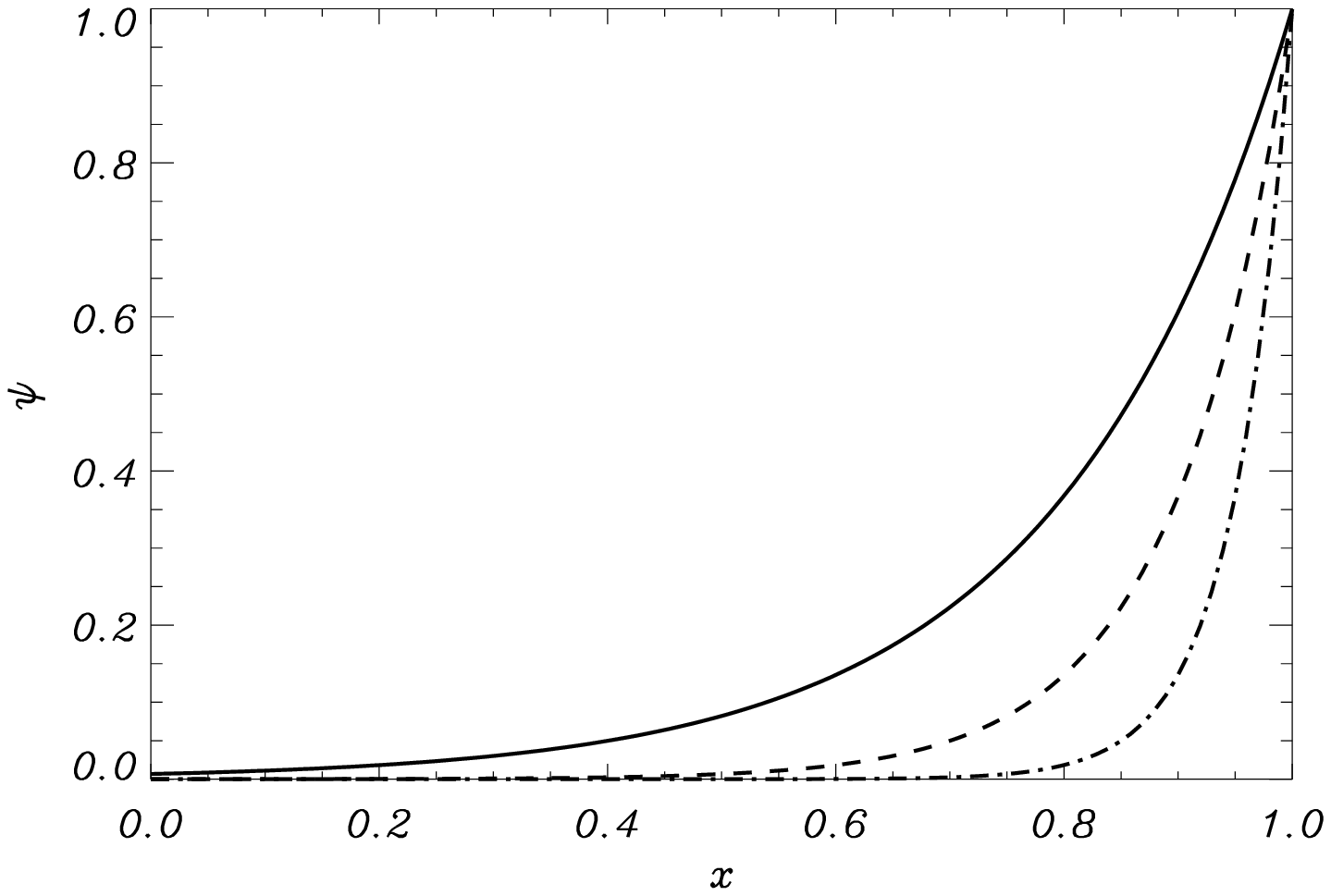}
\caption{The dependence of the toroidal field on the spherical radius
for $d=0.2$ (solid line), $0.1$ (dashed). and $0.05$ (dot-dashed).}
\end{center}
\end{figure}

\begin{figure}
\begin{center}
\includegraphics[width=10cm]{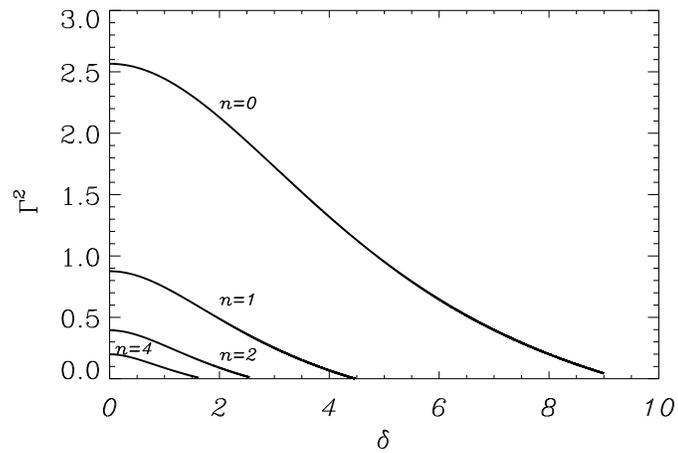}
\caption{The dimensionless growth rate as a function of $\delta$  for 
several eigenmodes with various number of nodes $n$ in the radial direction, 
for $m=1$, $l=20$, $d=0.1$ and $\varepsilon=0$.}
\end{center}
\end{figure}

\begin{figure}
\begin{center}
\includegraphics[width=10cm]{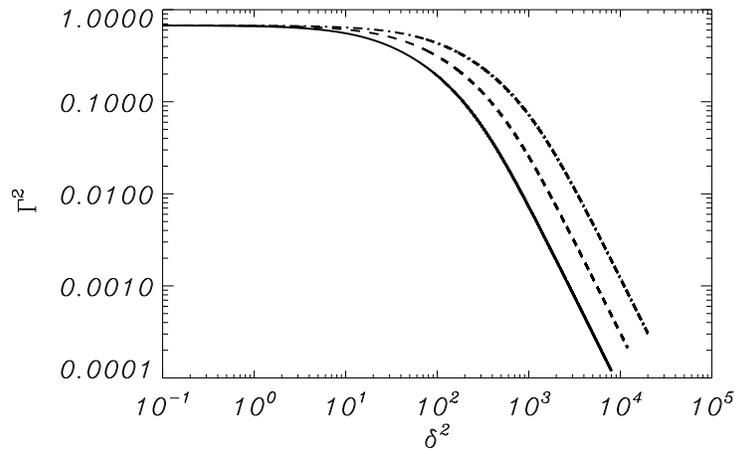}
\caption{The dimensionless growth rate as a function of $\delta^2$ for the
fundamental eigenmode with $m=1$, $l=20$, $d=0.1$, and three values of
$\varepsilon$: $10^{-2}$ (solid line), $2 \times 10^{-2}$ (dashed), and
$4 \times 10^{-2}$ (dot-dashed).}
\end{center}
\end{figure}

\end{document}